\documentclass[aps,showpacs]{revtex4-1}
\usepackage[dvips]{epsfig}
\usepackage[dvips]{color}
\usepackage{graphicx}
\usepackage{float}
\usepackage{subfigure}
\usepackage[applemac]{inputenc}
\usepackage[T1]{fontenc}
\usepackage{amssymb}
\usepackage{amsmath}
\usepackage{bm}
\usepackage{mathrsfs}
\usepackage{float}
\usepackage{slashed}
\usepackage[english]{babel}
\usepackage{comment}

\begin{document}

\title{Linear and nonlinear wave propagation in weakly relativistic quantum
plasmas}
\author{Martin Stefan}
\email{martin.stefan@physics.umu.se}
\author{Gert Brodin}
\email{gert.brodin@physics.umu.se}
\affiliation{Department of Physics, Ume{\aa } University, SE--901 87
Ume{\aa}, Sweden}

\begin{abstract}
We consider a recently derived kinetic model for weakly relativistic quantum
plasmas. We find that that the effects of spin-orbit interaction and Thomas
precession may alter the linear dispersion relation for a magnetized plasma
in case of high plasma densities and/or strong magnetic fields. Furthermore,
the ponderomotive force induced by an electromagnetic pulse is studied for
an unmagnetized plasma. It turns out that for this case the spin-orbit
interaction always give a significant contribution to the quantum part of
the ponderomotive force.
\end{abstract}

\maketitle

\section{Introduction}

Much recent work have studied quantum effects in plasmas. Applications
typically occur for high density plasmas of low or moderate temperature, see
e.g. Refs \cite{Manfredi,glenzer-redmer,Shukla-Eliasson,Shukla-Eliasson-RMP}
for detailed discussions. Typically hydrodynamic approaches cover effects
due to particle dispersion and the Fermi pressure \cite{Haas2005,SW-19},
whereas much of the kinetic treatments is based on the Wigner equation \cite%
{Manfredi}. More accurate treatments based on the Kadanoff-Baym kinetic
equations \cite{Kadanoff-Baym,Bonitz -1998} are also common. \ Moreover, the
magnetization currents and magnetic dipole force due to the electron spin
have been included in hydrodynamic \cite{zamanian2010fluid} as well as
kinetic theories \cite{zamanian2010kin}.

Relativistic effects may also have considerable influence over the dynamics 
\cite{NL-effects}, due to e.g high power lasers \cite{Nonlinear evolution}.
Combined quantum mechanical and relativistic effects in plasmas might be
possible to probe with soon to be built lasers, in particular x-ray lasers
like X-FEL (e.g \cite{XFEL-DESY}) and future lasers with even higher photon
energies. Also in astrophysics there are domains where quantum relativistic
plasma dynamics are of importance \cite{Astro-Drake}.

In the present paper we will study wave propagation using a recently
developed kinetic model \cite{asenjo}, extending previous quantum mechanical
models \cite{zamanian2010kin}, to including weakly relativistic effects such
as spin-orbit interaction and Thomas precession \cite{Jackson-Thomas}.
Firstly the dispersion relation for electromagnetic waves propagating
parallell to an external magnetic field is derived. It is found that the new
terms modify the dispersion relation for high densities and/or strong
magnetic fields. The linear investigations is then taken as a starting point
for calculating the ponderomotive force due to electromagnetic pulses. The
classical version of this problem has been thoroughly studied in the past
decades \cite{Karpman-pond}, and the ponderomitive force is known to give
raise to effects such as wakefield generation \cite{Wakefield}, soliton
formation, self-focusing and wave collapse \cite{NL-effects}. The
contributions to the ponderomotive force from non-relativistic spin dynamics
have been studied by Refs. \cite{Spin-pond-1,stefanpond}, and the it has
been shown that the this may lead to spin-polarization of an initially
unpolarized plasma. For the specific case of an unmagnetized plasma we find
that the weakly relativistic effects significantly modify the spin
contribution to the ponderomotive force. Furthemore, the spin quantum
contribution becomes comparable to the classical ponderomitive force when
the electromagnetic wavelength approaches the Compton wavelength. The
implications of are results are discussed.

\section{Basic model and linear solution}

In a recent work Asenjo et al \cite{asenjo} presented a kinetic evolution
equation for a weakly relativistic spin 1/2 collisionless plasma in the long
scale lenght limit, to order $c^{-2}$ and $h^2$: 
\begin{align}
0 &= \frac{\partial f}{\partial t} + \left\{ \frac{ \bm p }{ m } + \frac{
\mu }{ 2 m c } \bm E \times ( \bm s + \nabla_s ) \right\} \cdot \nabla_x f
+q \left( \frac{1}{c}\left\{ \frac{ \bm p }{ m } + \frac{ \mu }{ 2 m c } \bm %
E \times ( \bm s + \nabla_s ) \right\} \times \bm B + \bm E \right) \cdot
\nabla_p f  \notag \\
&+ \frac{2 \mu}{ \hbar} \bm s \times \left( \bm B - \frac{ \bm p \times \bm %
E }{ 2 m c } \right) \cdot \nabla_s f + \mu \left( \bm s + \nabla_s \right)
\cdot \left[ \partial_x^i \left( \bm B - \frac{\bm p \times \bm E }{ 2 m c}
\right) \right] \partial_p^i f ,  \label{1}
\end{align}
where $f = f(\bm x, \bm p, \bm s, t)$ is the quasi-distribution function
defined on a phase space extended by two spin dimensions (denoted $\bm s$)
on the unit sphere, in addition to the traditional space and momentum
coordinates $\bm x$ and $\bm p$. We use the notation $m$ for the mass and $%
\mu$ is the magnetic moment of the particle, $q$ is its charge. The index $x$%
, $p$ or $s$ on the nabla operator indicates that it acts on the respective
coordinates. When writing down this equation the last term of the equation
derived in \cite{asenjo} was omitted. This term is associated with the
Darwin term in the Hamiltonian, and was dropped since in a long wavelength
expansion it is smaller than the other terms.

This Vlasov-like equation is coupled to Maxwells equations 
\begin{align}
\nabla \cdot \bm E &= 4 \pi \rho_T , \\
\nabla \times \bm B &= \frac{1}{c} \frac{\partial \bm E}{\partial t } + 
\frac{4 \pi}{ c } \bm J_T,
\end{align}
where the total charge and current density are given by 
\begin{align}
\rho_T =& \rho_F + \nabla \cdot \bm P , \\
\bm J_T =& \bm J_F + \nabla \times \bm M + \frac{\partial}{\partial t } \bm %
P.
\end{align}
Here $\rho_F = q \int \! d\Omega \, f$ is the free charge density and the
free current density, the polarisation and magnetisation are given by 
\begin{align}
\bm J_f &= q \int \! d \Omega \, \left( \frac{\bm p}{ m } + \frac{3 \mu }{ 2
m c } \bm E \times \bm s \right) f, \\
\bm P &= - 3 \mu \int \! d \Omega \, \frac{\bm s \times \bm p}{ 2 m c } f, \\
\bm M &= 3 \mu \int \! d \Omega \, \bm s f.
\end{align}
We make the division $f = f_0 + \tilde f$ where $f_0$ is the background
distribution and $\tilde f$ is the perturbed distribution function, assumed
to be homogenious and isotropic in the momentum variable. To sum up our
model, it is a Vlasov-like equation for a quasi distribution function in a
phase space expanded by the spin variable $\bm s$, which is measured in the
rest frame of the particle. It contains the Lorentz force, magnetic dipole
force, Thomas correction and spin precession with spin orbit correction. It
should be noted that since this model is semirelativistic, the relation
between the momentum $\bm p$ and the kinetic velocity $\bm v$ is nontrivial.

After linearization, \eqref{1} reads 
\begin{align}
0& =\frac{\partial f_{1}}{\partial t}+\frac{\bm p}{m}\cdot \nabla
_{x}f_{1}+q\left( \frac{\bm p\times \bm B_{1}}{mc}+\left\{ \frac{\mu }{%
2mc^{2}}\bm E_{1}\times (\bm s+\nabla _{s})\right\} \times \hat{z}B_{0}+\bm %
E_{1}\right) \cdot \nabla _{p}f_{0}+\frac{q}{m^{2}}\bm p\times \hat{z}%
B_{0}\cdot \nabla _{p}f_{1}  \notag \\
& +\frac{2\mu }{\hbar }\bm s\times \left( \bm B_{1}-\frac{\bm p\times \bm %
E_{1}}{2mc}\right) \cdot \nabla _{s}f_{0}+\frac{2\mu B_{0}}{\hbar }\bm %
s\times \hat{z}\cdot \nabla _{s}f_{1}+\mu \left( \bm s+\nabla _{s}\right)
\cdot \left[ \partial _{x}^{i}\left( \bm B_{1}-\frac{\bm p\times \bm E_{1}}{%
2mc}\right) \right] \partial _{p}^{i}f_{0}  \notag \\
& -\frac{\hbar ^{2}q}{8m^{2}c^{2}}\left[ \partial _{x}^{i}(\nabla \cdot \bm %
E).\right] \partial _{p}^{i}f_{0}  \label{linearised}
\end{align}%
We use a standard ansatz of quasi-monochromatic harmonic variation on the
perturbed quantities, $\bm E_{1}=\tilde{\bm E_{1}}\exp i(\bm k\cdot \bm %
x-\omega t)$ etc, and choose $\bm k=k\hat{z}$ and the polarization in the $%
(x,y)$-plane. The external magnetic field $\bm B_{0}$ is assumed to be
static, homogeneous and point in the $\hat{z}$-direction. For the momentum
variable we use cylindrical coordinates, and for the spin we use spherical
coordinates on the unit sphere. Furthermore the unperturbed spins are
assumed to be in thermal equilibrium, and thus $f_{0}(\bm s,p)=f_{0}(p)\left[
1+\tanh \left( \mu B_{0}/k_{B}T\right) \cos \theta _{s}\right] $ \cite%
{Thermo-note}.

Now we expand $f_{1}$ in eigenfunctions 
\begin{equation}
f_{1}=(1/2\pi )\sum_{a,b=-\infty }^{\infty }W_{ab}(p_{\perp },p_{z},\bm %
x,\theta _{s},t)e^{-i(a\phi _{v}+b\phi _{s})}+\mathrm{c.c.},  \label{ansatz}
\end{equation}%
where $\mathrm{c.c.}$ stands for complex conjugate and insert (\ref{ansatz})
in \eqref{linearised}. Mutiplying with $1/(2\pi )\exp [i(n\phi _{p}+m\phi
_{s})]$ and integrating both these angles from $0$ to $2\pi $ we can solve
for $W_{a,b}$. Plugging this back into \eqref{ansatz} we get the disturbed
distribution function to first order in perturbed quantities as 
\begin{align}
f_{1}=& \sum_{\pm }\Bigg\{\frac{1}{2\pi i\left[ \omega -\frac{kp_{z}}{m}\pm
\omega _{c}\right] }\bigg[qp_{\perp }\pi E_{\pm }\left( \frac{-B_{0}\mu }{%
2mc^{2}}\left( \cos \theta _{s}-\sin \theta _{s}\frac{\partial }{\partial
\theta }_{s}\right) +1\right) 2\frac{\partial f_{0}}{\partial (p^{2})} 
\notag \\
& \mp \pi \mu k\frac{p_{\perp }E_{\pm }}{2mc}\left( \cos \theta _{s}-\sin
\theta _{s}\frac{\partial }{\partial \theta }_{s}\right) 2p_{z}\frac{%
\partial f_{0}}{\partial (p^{2})}\bigg]e^{\mp i\phi _{v}} \\
& +\frac{1}{2\pi i\left[ \omega -\frac{kp_{z}}{m}\pm \omega _{c}g\right] }%
\bigg[-\frac{\pi \mu }{\hbar }E_{\pm }\left( \frac{kc}{\omega }-\frac{p_{z}}{%
2mc}\right) \frac{\partial }{\partial \theta }_{s}f_{0}  \notag \\
& \mp \frac{\pi \mu k}{2}\left( \frac{kc}{\omega }-\frac{p_{z}}{2mc}\right)
E_{\pm }\left( \sin \theta _{s}+\cos \theta _{s}\frac{\partial }{\partial
\theta }_{s}\right) 2p_{z}\frac{\partial f_{0}}{\partial (p^{2})}\bigg]%
e^{\mp i\phi _{s}}\Bigg\}+\mathrm{c.c},  \notag
\end{align}%
where $\omega _{c}=qB_{0}/mc$ is the cyclotron frequency, $\omega
_{cg}=(g/2)\omega _{c}$ is the spin precession frequency and $E_{\pm }\equiv
E_{x}\pm iE_{y}$.

In combination with Maxwell's equations and the expressions for the
currents, we obtain the dispersion relation 
\begin{equation}
\omega^2 - k^2 c^2 = 4 \pi \omega \left[ 2 \int \! d^3 p \, (\alpha_\pm +
\beta_\pm) \mp \frac{q \mu}{ 2 m c } (n_{0+} - n_{0-}) \right]
\end{equation}
where 
\begin{align}
\alpha_\pm \equiv & \ - \pi \frac{ p_\perp^2 }{ \left[ \omega - \frac{k p_z}{%
m} \pm \omega_c \right] } \bigg\{ \left[ \left( 2 \frac{q}{m} \frac{ - B_0
\mu }{ 2 m c^2 } \mp \frac{ \mu \omega}{m c^2}\right) q - \frac{2 q}{m} 
\frac{ \pi \mu k}{2 m c } p_z\right] \frac{ \partial }{\partial (p^2)} \frac{%
1}{4 \pi} \left[ f_+(p^2) - f_-(p^2) \right]  \notag \\
&+ \left[ \left( 2 \frac{q}{m} \mp \frac{ 2 \mu \omega }{m c^2}\frac{ B_0
\mu }{ 2 m c^2 }\right) q \mp \frac{2 \mu }{mc}\frac{\pi \mu k}{2 mc}p_z %
\right] \frac{ \partial }{\partial (p^2)} \frac{1}{4 \pi} \left[ f_+(p^2) +
f_-(p^2) \right] \bigg\}  \notag \\
\beta_\pm \equiv & \ - \pi \, \frac{ 4 \mu }{2 \left[ \omega - \frac{k p_z}{m%
} \pm \omega_{cg} \right] } \left[ k \pm \frac{ \omega }{2 m c^2} p_z \right]
\bigg\{ - \frac{\mu}{ \hbar} \left( \frac{ k c }{\omega} - \frac{ p_z }{ 2 m
c } \right) \frac{1}{4 \pi} \left[f_+(p^2) - f_-(p^2) \right]  \notag \\
& \mp \frac{ \mu k}{2} \left( \frac{ k c }{ \omega } -\frac{p_z }{ 2 m c}
\right) 2 p_z \frac{ \partial f_0 }{\partial (p^2)} \frac{1}{4 \pi} \left[%
f_+(p^2) + f_-(p^2) \right] \bigg\} .
\end{align}
As expected the dispersion relation has two solutions,
corresponding to left and right circular polarisation.

Taking the long wavelenght limit $k \rightarrow 0$ the result simplifies to 
\begin{align}
\omega =& \ \frac{4 \pi }{c^2} \Bigg\{ \frac{ 1 }{ \left[ \omega \pm
\omega_c \right] } \bigg[ \left( \frac{q}{m} \frac{ - B_0 \mu }{ 2 m c^2 }
\mp \frac{ \mu \omega}{m c^2}\right) q( n_{0+} - n_{0-} ) + \left( \frac{q}{m%
} \pm \frac{ \mu \omega }{m c^2}\frac{ B_0 \mu }{ 2 m c^2 }\right) q (
n_{0+} + n_{0-} ) \bigg]  \notag \\
& \pm \frac{ \mu^2 p_t^2 }{ 2 \left[ \omega \pm \omega_{cg} \right] } \frac{
\omega }{ 4 \hbar m^2 c^3} \left[ n_{0+} - n_{0-} \right] \Bigg\} \mp \frac{%
2 \pi }{c^2} \frac{q \mu }{ m c } (n_{0+} - n_{0-}).
\end{align}
Here we can note that for high densities and low temperatures the terms
stemming from relativistic effects can significantly alter the dispersion
relation compared to previous nonrelativistic results like e.g. with \cite%
{lundin}. However we will now mainly be concerned with the case where the
impact of the new terms are small in a linear calculation, but might be of
importance when dealing with nonlinear problems.

\section{The ponderomotive force}

The classical pondermotive force has been thoroughly studied, and recently
pure spin effects have also been explored. In the present paper we are
mostly concerned with the effects arising from spin-orbit coupling, which
can be considered as a quantum relativistic effect. To see these effects, it
suffices to study an unmagnetised plasma, thus also reducing the algebra to
more manageable proportions. This also allows us to make the very reasonable
assumption that the equilibrium distribution function does not depend on $%
\theta _{s}$. This can be motivated since we do not have any external field,
so the spin states are degenerate. This reduces the first order distribution
function to 
\begin{align}
f_{1}=& \sum_{\pm }\frac{1}{2\pi i\left[ \omega -\frac{kp_{z}}{m}\right] }%
\Bigg\{\bigg[qp_{\perp }\pi E_{\pm }2\frac{\partial f_{0}}{\partial (p^{2})}%
\mp \pi \mu k\frac{p_{\perp }E_{\pm }}{2mc}\cos \theta _{s}2p_{z}\frac{%
\partial f_{0}}{\partial (p^{2})}\bigg]e^{\mp i\phi _{p}}  \notag \\
& \mp \frac{\pi \mu k}{2}\left( \frac{kc}{\omega }-\frac{p_{z}}{2mc}\right)
E_{\pm }\sin \theta _{s}2p_{z}\frac{\partial f_{0}}{\partial (p^{2})}e^{\mp
i\phi _{s}}\Bigg\}+\mathrm{c.c}.  \label{f-2}
\end{align}%
Now we study the evolution equation to second order in perturbed quantities,
and only keep source terms on the low frequency time scale thus obtaining 
\begin{align}
\frac{\partial }{\partial t}f_{lf}+\frac{\bm p}{m}\cdot \nabla _{x}f_{lf}=& -%
\frac{\mu }{2mc}\bm E_{1}\times (\bm s+\nabla _{s})\cdot \nabla
_{x}f_{1}^{\ast }-\frac{q\mu }{2mc^{2}}\left[ \bm E_{1}\times (\bm s+\nabla
_{s})\right] \times \bm B_{1}^{\ast }\cdot \nabla _{p}f_{0}  \notag \\
& -\frac{2\mu }{\hbar }\bm s\times \left( \bm B_{1}-\frac{\bm p\times \bm %
E_{1}}{2mc}\right) \cdot \nabla _{s}f_{1}^{\ast }-\mu (\bm s+\nabla
_{s})\cdot \left[ \partial _{x}^{i}\left( \bm B_{1}-\frac{\bm p\times \bm E}{%
2mc}\right) \right] \partial _{p}^{i}f_{1}^{\ast }+\mathrm{c.c},
\label{vlasovnl}
\end{align}%
where the star denotes complex conjugate and the index \textit{lf} indicates
that the quantity has no rapid oscillations, and thus varies only on the slow
spatial and temporal scales. What we want to calculate is the total low
frequency current, composed in our case by the free current 
\begin{equation}
J_{flf}=q\int \!d\Omega \,\left( \frac{p_{z}}{m}f_{lf}+\frac{3\mu }{2mc}\bm %
E_{1}\times \bm sf_{1}^{\ast }+\frac{3\mu }{2mc}\bm E_{1}^{\ast }\times \bm %
sf_{1}\right) ,
\end{equation}%
and the polarisation current 
\begin{equation}
J_{Plf}=-3\mu \frac{\partial }{\partial t}\int \!d\Omega \,\frac{p_{\perp
}\sin \theta _{s}(\cos \phi _{s}\sin \phi _{p}-\sin \phi _{s}\cos \phi _{p})%
}{2mc}f_{lf}.
\end{equation}%
We note that the magnetisation current vanishes in our geometry. For
simplicity we only consider the quantum contributions, since the classical
ponderomotive current has been calculated in a number of previous works
already. Starting with the free current we note that 
\begin{equation}
\frac{\partial }{\partial t}J_{flf}+q\int \!d\Omega \frac{p_{z}^{2}}{m^{2}}%
\frac{\partial }{\partial z}f_{lf}=q\int \!d\Omega \frac{p_{z}}{m}\left( 
\frac{\partial }{\partial t}+\frac{p_{z}}{m}\frac{\partial }{\partial z}%
\right) f_{lf}+\frac{\partial }{\partial t}q\int \!d\Omega \frac{3\mu }{2mc}(%
\bm E\times \bm s)_{z}f_{1}^{\ast }.
\end{equation}%
Using \eqref{vlasovnl} in combination with \eqref{f-2} the we can calculate
the first term on the right hand side in terms of the field. The integral on
the left hand side can be dealt with by noting that the convective
derivative in the evolution equation for the low frequency distribution
function is small in the low temperature limit, and can thus be calculated
using perturbation theory. Now we obtain an expression for the second time
derivative of the free current: 
\begin{equation}
\frac{\partial ^{2}}{\partial t^{2}}J_{flf}\simeq q\frac{\partial }{\partial
t}\int \!d\Omega \frac{p_{z}}{m}\left( \frac{\partial }{\partial t}+\frac{%
p_{z}}{m}\frac{\partial }{\partial z}\right) f_{lf}+\frac{\partial ^{2}}{%
\partial t^{2}}q\int \!d\Omega \frac{3\mu }{2mc}(\bm E\times \bm %
s)_{z}f_{1}^{\ast }+q\int \!d\Omega \frac{p_{z}^{2}}{m^{2}}\frac{\partial }{%
\partial z}\left( \frac{\partial }{\partial t}+\frac{p_{z}}{m}\frac{\partial 
}{\partial z}\right) f_{lf}.  \label{Eq-free-current-2}
\end{equation}%
The approximation performed to obtain (\ref{Eq-free-current-2}) is the
addition of the last term proportional to $(p_{z}/m)\partial /\partial z$.
This addition is a higher order thermal correction, but is useful since it
enable us to rewrite the terms involving $f_{lf}$ by using Eq (\ref{vlasovnl}%
) combined with (\ref{f-2}) to obtain a driving term for the low-frequency
current proportional to the high-frequency wave intensity (i.e. proportional
to $\left\vert E_{\pm }\right\vert ^{2}=$ $\left\vert E_{x}\right\vert
^{2}+\left\vert E_{y}\right\vert ^{2}$) Following the same approximate
procedure for the polarisation current we have 
\begin{align}
\frac{\partial ^{2}}{\partial t^{2}}J_{plf}& \simeq \ -3\mu \frac{\partial
^{2}}{\partial t^{2}}\int \!d\Omega \,\frac{p_{\perp }\sin \theta _{s}(\cos
\phi _{s}\sin \phi _{p}-\sin \phi _{s}\cos \phi _{p})}{2mc}\left( \frac{%
\partial }{\partial t}+\frac{p_{z}}{m}\frac{\partial }{\partial z}\right)
f_{lf}  \notag \\
& + 3 \mu \frac{\partial }{\partial t}\int \!d\Omega \,\frac{p_{\perp }\sin
\theta _{s}(\cos \phi _{s}\sin \phi _{p}-\sin \phi _{s}\cos \phi _{p})}{2mc}%
\frac{p_{z}}{m}\frac{\partial }{\partial z}\left( \frac{\partial }{\partial t%
}+\frac{p_{z}}{m}\frac{\partial }{\partial z}\right) f_{lf}.
\label{Eq-pol-current}
\end{align}%
Using again eqs (\ref{vlasovnl}) and (\ref{f-2}) to to rewrite the right hand
source terms and combining the results from (\ref{Eq-free-current-2}) and (%
\ref{Eq-pol-current}) we find that the second order time derivative for the
total current is given by 
\begin{align}
\frac{\partial ^{2}}{\partial t^{2}}J=-& \frac{8}{3}q\frac{\pi \mu ^{2}k^{2}%
}{m^{2}\omega ^{2}}\Bigg\{\left[ \frac{11}{2}\left( 1-\frac{k^{2}c^{2}}{%
\omega ^{2}}\right) \frac{\partial }{\partial z}+\frac{1}{2c}\left( 1+\frac{%
kc}{\omega }-\frac{2\omega }{kc}\right) \frac{\partial }{\partial t}-\frac{%
4k^{2}c^{2}}{\omega ^{2}}\frac{\partial }{\partial z}-\frac{2k^{2}c^{2}}{%
\omega ^{2}}\left( \frac{\partial }{\partial z}+2\frac{k}{\omega }\frac{%
\partial }{\partial t}\right) \right.   \notag \\
& +\left. \frac{3}{2}\frac{\omega }{c^{2}k}\left( 1+\frac{k^{3}c^{3}}{\omega
^{3}}\right) \frac{\partial }{\partial t}\right] \frac{\partial }{\partial t}%
+3\frac{kc^{2}}{\omega }\frac{\partial ^{2}}{\partial z^{2}}\Bigg\}%
(|E_{x}|^{2}+|E_{y}|^{2})n_{0},
\end{align}%
where we have expanded each term to lowest order in $p_{z}$ to be able to
perform the integration (which is consistent with the approximations in (\ref%
{Eq-free-current-2}) and (\ref{Eq-pol-current}) ), and defined the plasma
frequency $\omega _{p}^{2}=4\pi q^{2}n_{0}/m$. The classical contribution
has been omitted for simplicity. Using the time derivative of Ampere's law
we obtain 
\begin{align}
\frac{\partial }{\partial t}\left( \frac{\partial ^{2}}{\partial t^{2}}%
+\omega _{p}^{2}\right) E_{lf}=& \frac{32}{3}q\frac{\pi ^{2}\mu ^{2}k^{2}}{%
m^{2}\omega ^{2}}\Bigg\{\left[ \frac{11}{2}\left( 1-\frac{k^{2}c^{2}}{\omega
^{2}}\right) \frac{\partial }{\partial z}+\frac{1}{2c}\left( 1+\frac{kc}{%
\omega }-\frac{2\omega }{kc}\right) \frac{\partial }{\partial t}-\frac{%
4k^{2}c^{2}}{\omega ^{2}}\frac{\partial }{\partial z}-\frac{2k^{2}c^{2}}{%
\omega ^{2}}\left( \frac{\partial }{\partial z}+2\frac{k}{\omega }\frac{%
\partial }{\partial t}\right) \right.   \notag \\
& +\left. \frac{3}{2}\frac{\omega }{c^{2}k}\left( 1+\frac{k^{3}c^{3}}{\omega
^{3}}\right) \frac{\partial }{\partial t}\right] \frac{\partial }{\partial t}%
+3\frac{kc^{2}}{\omega }\frac{\partial ^{2}}{\partial z^{2}}\Bigg\}%
(|E_{x}|^{2}+|E_{y}|^{2})n_{0}.
\end{align}%
We can note that the second to last term in the square bracket is what was
obtained in previous works based on models not containing relativistic
effects \cite{stefanpond}. If we assume that $kc/\omega $ is roughly of
order $1$ we see that all terms in the square brackets are of the same
order. This implies that when dealing with an unmagnetised plasma where spin
effects are important, the spin orbit coupling contributions must be taken
into account as well. Furthermore, in the approximation where $\omega \gg
\omega _{p}$ such that $\partial /\partial t=c\,\partial /\partial z$ we
have 
\begin{equation}
\left( \frac{\partial ^{2}}{\partial t^{2}}+\omega _{p}^{2}\right)
E_{lf}=-8\pi \frac{8}{3}q\frac{\pi \mu ^{2}k^{2}}{m^{2}\omega ^{2}}\frac{%
\partial }{\partial z}(|E_{x}|^{2}+|E_{y}|^{2})n_{0}.
\end{equation}%
This spin contribution should be compared with the classical current given
by 
\begin{equation}
\left( \frac{\partial ^{2}}{\partial t^{2}}+\omega _{p}^{2}\right) E_{lfc}=%
\frac{q\omega _{p}^{2}}{8m\omega ^{2}}\frac{\partial }{\partial z}\left(
|E_{x}|^{2}+|E_{y}|^{2}\right) ,
\end{equation}%
and we see that these two source terms will be comparable if $\hbar k/mc\sim
1$. For typical parameters where $\omega \sim kc$ this implies that we need
photon energies of the order of the electron rest mass energy, i.e. gamma
rays. Here it should be stressed that for such short wavelengths several
other effects that have been omitted is likely to be of importance, for
example particle dispersive effects and the Darwin term \cite{asenjo}.

\section{Summary and conclusions}

In this work we have first solved the linear problem and found the
dispersion relation for waves propagating parallel to the external magnetic
field in a weakly relativistic spin plasma, and seen that it gives the
correct classical limit. It was also seen that in the long wavelength limit
the dispersion relation is significantly modified by the spin-orbit
interaction when the Zeeman energy approaches the electron rest mass energy.

Furthermore the nonlinear ponderomotive force in an unmagnetized setting has
been derived, and compared with previous classical and nonrelativistic
quantum result. In the case of high energy radiation the quantum
contributions are seen to actually dominate over the classical ones. For
example if a plasma is illuminated by gamma rays in the $\mathrm{MeV}$%
-regime. The planned high power x-ray lasers like Xfel promise photon
energies around $25\mathrm{keV}$, \cite{XFEL-DESY} which is still not enough
for the quantum terms to dominate. However, accelerating an electron bunch
with a lineac might suffice to blue-shift the x-ray photons enough to give
them $\mathrm{MeV}$-energies in the rest frame of the electrons.

Another possibility to obtain the energetic photons is from Gamma ray bursts
, see eg \cite{piran} for a review. In the GRB itself the radiation is
created in a relativistic jet, which means that the photons are redshifted
in the reference frame of the plasma. However there is a possibility that
the radiation passes through an accretion disc between the source and the
observer, and when this happens the ponderomotive forces and subsequent
acceleration of the particles in the accretion disc can be dominated by the
terms calculated above.

When dealing with highly energetic photons we always face the possibility of
QED effects, which are not included in the current model. To start with we
consider Compton scattering. Studying the Klein-Nishina cross section \cite%
{jackson-2} we see that the cross section actually decreases with the photon
energy for energetic photons, thus this mechanism will be suppressed. Pair
production on the other hand can play a dominating part if photon energies
are high, but in order to conserve energy and momentum the two interacting
photons need to be of different energy. If we only consider
quasi-monochromatic beams or beams with a narrow energy spectrum the photons
will therefore not produce pairs, and we conclude that it is consistent in
this case to neglect QED effects. Furthermore, the particle dispersive
effects neglected in the kinetic model may also play a role for high photon
energies.

\end{document}